\documentstyle[preprint,aps,prl]{revtex}
\begin{document}
\draft
\title{
The Feynman effective classical potential
in the Schr\"odinger formulation. 
}
\author{Rafael Ram\'{\i}rez , Telesforo L\'opez-Ciudad, and Jos\'e C. Noya} 
\address{Instituto de Ciencia de Materiales,
         Consejo Superior de Investigaciones Cient\'{\i}ficas
         (C.S.I.C.),
         Cantoblanco, 28049 Madrid, Spain   }
\date{\today}
\maketitle
\begin{abstract}
New physical insight into 
the correspondence between path integral concepts
and the Schr\"odinger formulation is gained by the analysis of
the effective classical potential, that is defined within the 
Feynman path integral formulation of statistical mechanics.  
This potential is related to   
the quasi-static response of the equilibrium system to an external force.
These findings allow for a comprehensive formulation of
dynamical approximations based on this 
potential.\\
\end{abstract}
\pacs{PACS numbers: 05.30.-d, 03.65.Ca, 82.20.Wt, 82.20.Db}

The path integral formulation of statistical
mechanics is a powerful method to study
quantum many-body systems.
An essential property of this approach is the mapping
of a quantum system onto a classical model of
harmonic ring polymers, 
whose equilibrium properties can be
studied with high accuracy. \cite{ceperley95} However, 
dynamical properties at finite
temperatures can not be derived with the same sort of rigor,
as the solution of the real time path integrals involves
complex valued functionals without a suitable sampling function
for stochastic integration.

The {\it effective classical potential} (ECP), \cite{cuccoli95}
that was introduced by Feynman to study 
systems in thermodynamic
equilibrium, \cite{feynman65} has been 
a central quantity to derive two quantum dynamical 
approximations, the {\it quantum transition state theory} (QTST),
\cite{gillan87} that
aims at calculating rate constants of activated processes, and
the {\it centroid molecular dynamics} (CMD), \cite{cao94} that
aims at calculating real time correlation functions for 
quantum particles. 
The definition of the ECP is based on
the concept of {\it path centroid}, 
that is the center of gravity of each of the 
ring polymers that represent the quantum particle. 
However, in spite of the extensive use of the path centroid in condensed
matter and chemical physics studies, its relation 
to a measurable physical observable and hence 
the physical meaning of the derived dynamical approximations
remain largely unexplained.

In this work, we aim at a deeper understanding of
the equivalence between the
path integral and the Schr\"odinger formulation by 
showing that the centroid coordinate is 
related to the quasi-static response of the 
system to an external force.
The Schr\"odinger formulation provides
new physical insight into the dynamical
approximations based on the ECP.
At zero temperature, the ECP is
the mean energy of {\it minimum energy  
wave packets} (MEWP's, to be defined below) 
and the CMD is an approximate dynamics
based on these wave packets.
Technical details of the analysis are left for  
a later work. \cite{ramirez}

We begin by reviewing the definition of the ECP 
in the path integral formulation.
The simple case 
of a quantum particle of mass $m$ 
having bound states in a one-dimensional potential
$V(x)$ is considered. 
The Hamiltonian of the particle is $H_0$. 
The extension to the many-particle case is, for distinguishable
particles, straightforward. \cite{ramirez} 
At a given temperature the equilibrium properties
are derived from the partition function, $Z_0$, and from the
particle's probability density, $\rho(x)$. Both are related by
the expression $Z_0=\int_{-\infty}^{\infty} dx \rho(x)$. The 
path integral formulation of $\rho(x)$ is
\cite{feynman65}

\begin{equation}
\rho(x) = \int_{x=x(0)}^{x=x(\beta\hbar)} D[x(u)] 
             \exp \left( - \frac{S[x(u)]}{\hbar} \right)   \, ,
\label{rhox}
\end{equation}
where $S[x(u)]$ is the functional of the Euclidean action of the
path $x(u)$, $u$ is the imaginary time that varies between 0
and $\beta\hbar$, and $\beta$ is 
the inverse temperature $(k_BT)^{-1}$. 
The paths $x(u)$ can be considered
as alternative ways for the propagation of the particle,
and the sum over paths can
be analyzed in many different ways depending on the different
classes into which the alternatives can be divided. \cite{feynman65}
The way conducting to the ECP
uses the centroid or average point, $x_c$, of the path $x(u)$ 

\begin{equation}
x_c =  \frac{1}{\beta\hbar}\int_0^{\beta\hbar} du \,\, x(u)    \, .
\label{xc}
\end{equation}
A class of paths is the subset of paths that have the same centroid.
A constrained path integral over the class of paths  
with centroid at $X$ is defined by introducing a delta function
in the integrand of Eq. (\ref{rhox}),

\begin{equation}
\rho_X(x) = 
      \int_x^x D[x(u)] \delta(X-x_c) 
             \exp \left( - \frac{S[x(u)]}{\hbar} \right)   \, .
\label{rhoXx}
\end{equation}
$\rho_X(x)$ may be considered
as a probability density around the centroid position $X$.
As illustration, the normalized function 
$Z(X)^{-1}\rho_X(x)$ obtained by a Monte
Carlo path integral simulation of a particle of mass $m=16$ au
in a double-well potential,
$V_{dp}(x) = \frac{1}{4}(x^2-1)^2$, is represented in Fig. \ref{f1}. 
The probability density is shown for four values of the 
centroid coordinate $X$. We also show the probability density
of the ground state of the potential $V_{dp}(x) - f x$, for 
several values of the parameter $f$, that represents an
external force acting on the particle. The identity between
these ground state probability densities and $Z(X)^{-1}\rho_X(x)$ 
will be explained below.
The normalization constant $Z(X)$
is an important quantity

\begin{equation}
Z(X) = \int_{-\infty}^{\infty} dx \rho_X(x)   \, ,
\label{ZX1}
\end{equation}
which has the physical meaning of a probability density for
the class of paths with centroid at $X$ and it is called
the {\it centroid density}. \cite{cao94}
The partition function can be recovered by a sum over all classes, i.e.,

\begin{equation}
Z_0 = \int_{-\infty}^{\infty} dX Z(X)   \, .
\label{Z0}
\end{equation}
This integral,
after substitution of $Z(X)$ by the following definition,

\begin{equation}
       Z(X) = 
      \left( \frac{m}{2 \pi \beta \hbar^2 } \right)^{1/2}        
             \exp \left[ - \beta F_{ef}(X) \right]         \, ,
\label{ZX2}
\end{equation}
has the same form
as the partition function of a classical particle moving in the
potential, $F_{ef}(X)$, which is the ECP. 
We summarize some properties of the ECP: \cite{cuccoli95,acocella95} 
{\it i)} all the thermodynamic properties that depend on $Z_0$
can be derived from it; 
{\it ii)} it is temperature dependent; 
{\it iii)} its calculation is      
analytical only for quadratic potentials, but variational
approximations are available;
{\it iv)} its high temperature limit is the
actual potential $V(x)$.

    
The centroid density $Z(X)$ (or the ECP) and the 
probability densities $Z(X)^{-1}\rho_X(x)$ 
are important quantities in the theory of path integrals, 
whose correspondence to the Schr\"odinger
formulation has not been clearly stated. 
In the following, we show how 
the centroid density is related to a physical observable.
We first consider                                      
a Hamiltonian depending on an external force $f$
acting on the particle,
$H_f = H_0 - fx$.
The path integral representation of the partition
function $Z_f$, corresponding to the Hamiltonian
$H_f$, can be expressed as\cite{kleinert93}   

\begin{equation}
 Z_f  = \int_{-\infty}^\infty dX  Z(X) e^{\beta f X}  
      = Z_0  \langle e^{\beta f X}  \rangle    \, ,
\label{Zf}
\end{equation}
where $Z(X)$ is the centroid density 
for the particle with Hamiltonian $H_0$.  
The angle brackets show an
average over the normalized centroid density. 
If $Z(X)$ is known, the partition
function $Z_f$ can be derived by Eq. (\ref{Zf})
for any arbitrary value of the external force $f$.
Thus, the analysis of classes of paths with fixed centroid 
allows to derive
the thermodynamic properties of a whole family of systems
whose Hamiltonian depends on a parameter $f$.
The moments of the centroid density are defined as

\begin{equation}
  \langle X^n \rangle = Z_0^{-1} \int_{-\infty}^\infty dX Z(X) X^n \, .
\label{Xn}
\end{equation}

The physical meaning of 
Eq. (\ref{Zf}) 
is that the {\it ratio of partition 
functions $Z_f/Z_0$ is the function generating the moments
of the centroid density}.
The moments 
are derived by differentiation of $Z_f/Z_0$ 
with respect to the variable $\beta f$. 
The result is simpler if we use the free energy, $F_f$,
defined as
$\exp(-\beta F_f)=Z_f$. 
The first two moments are obtained as

\begin{equation}
 \langle X \rangle = 
  - \left( \frac{\partial F_f }{\partial f} \right)_{f=0} \, ,
\label{X1}
\end{equation}
\begin{equation}
 \langle X^2 \rangle - \langle X \rangle^2 
= -  k_B T  \left( \frac{\partial^2 F_f }{\partial f^2} \right)_{f=0}
                                                           \, .
\label{X2}
\end{equation}
The mean-squared deviation of $X$ 
is often called the classical
delocalization of the particle.
Higher order derivatives of $F_f$ lead to higher moments
of the centroid density. \cite{ramirez} 
If the moments are known, 
then the centroid density itself will be fully determined.
We have then a formal relation between {\it the
centroid density and the change in the
free energy as a function of a quasi-static
force acting on the particle.}
This central result reveals   
the physical meaning of the centroid density 
within the Schr\"odinger formulation.

This correspondence between the centroid density 
and the Schr\"odinger formulation is valid for 
arbitrary temperatures. 
The zero temperature limit is particularly interesting,
as quantum effects
are then most important.
Before analyzing this limit,
it is convenient to study a property
of the Hamiltonian $H_f$
that leads to the definition
of MEWP's.
We look for quantum
states of the particle whose mean energy, 
$E^{min}_0=\langle \psi_f | H_0 | \psi_f\rangle$, is minimum
against small variations of $|\psi_f\rangle$.
Moreover, the states must satisfy two constraints:
{\it i)} their mean position is fixed at an arbitrary value
$\bar{x}=\langle \psi_f | x | \psi_f\rangle$; 
{\it ii)} they are normalized $\langle \psi_f | \psi_f\rangle=1$.
By straightforward application of calculus of variations, \cite{ramirez}
one finds that $ | \psi_f\rangle$ is the ground state of $H_f$

\begin{equation}
(H_0 - f x) |\psi_f\rangle = E_f |\psi_f\rangle       \, ,
\label{Hf}
\end{equation}
$f$ and $E_f$ are Lagrange multipliers
chosen so that the constraints {\it i} and {\it ii}
are satisfied. 
Note that $f$ is an implicit function  
of the arbitrary position $\bar{x}$, i.e., $f \equiv f(\bar{x})$.
The minimum energy, $E_0^{min}$, as a function of 
$\bar{x}$, is derived from Eq. (\ref{Hf}) as

\begin{equation}
 E^{min}_0(\bar{x}) = E_f + f \bar{x}               \, .
\label{Emin}
\end{equation}
We call the states $|\psi_f\rangle$ the
MEWP's of the potential $V(x)$
whose mean energy is $E^{min}_0(\bar{x})$. 
We show next how these states
are related to the ECP at zero temperature. 

At $T=0$, the free energy is equal to
the ground state energy,
$F_f = E_f$, and from Eqs. (\ref{X1}),
and (\ref{X2}) we obtain 

\begin{equation}
 \langle X \rangle = -\left( \frac{\partial E_f } 
                                  {\partial f} \right)_{f=0}  \,\,;\,\,
 \langle X^2 \rangle - \langle X \rangle^2 = 0 \, .
\label{XXX}
\end{equation}
$\langle X \rangle$ is then the
expectation value of
the derivative of the Hamiltonian operator in Eq. (\ref{Hf})
with respect to $f$ at $f=0$, i.e., $\langle X \rangle$
is equal to the mean position of the ground state of the
Hamiltonian $H_0$

\begin{equation}
 \langle X \rangle = \langle \psi_0 | x | \psi_0 \rangle = \bar{x}_0 \, .
\label{X}
\end{equation}
The mean-squared deviation of the centroid density vanishes
[Eq. (\ref{XXX}], 
thus the centroid density must be a delta function centered at $\bar{x}_0$. 
This result implies that 
only the class of paths with centroid at $X \equiv \bar{x}_0$ contributes
to the path integral in Eq. (\ref{rhox}), i.e.,            

\begin{equation}
 \lim_{T\to 0} Z_0^{-1} \rho(x)      
 = \lim_{T\to 0} Z(X)^{-1} \rho_{X}(x) \,\,\, , \,\,\, 
   \text{for}  \,\,\,  X  \equiv  \bar{x}_0     \, .
\label{limit}
\end{equation}
This identity provides
a clear physical interpretation of path integral quantities at
$T=0$. It implies that the centroid density 
and the partition function of the particle
are related by: $Z(X) = Z_0 \; \delta(X-\bar{x_0})$. 
Moreover, the asymptotic behavior of $Z_0$ and $Z(\bar{x}_0)$ as
$T \to 0$ is given by exponential functions of the ground
state energy, $\exp( - \beta E_0)$, and the ECP, 
$ \exp[ - \beta F_{ef}(\bar{x}_0)]$,
respectively. Then, $F_{ef}(\bar{x}_0)$
must be equal to the ground state energy.
This result is a particular case of a 
relation valid for any arbitrary centroid position $X$.
An essential step for this generalization is that the 
probability densities
$Z(X)^{-1} \rho_X(x)$ are invariant 
under any change in the Hamiltonian $H_0$ that is linear
in the coordinate $x$. 
The final result of this generalization is: \cite{ramirez}
{\it i)}
{\it at $T=0$                  
the normalized probability density 
for a given class of paths,
$Z(X)^{-1} \rho_X(x)$,
is identical to the  
probability density
of the MEWP, 
$| \langle x | \psi_f \rangle |^2$,
whose mean position is at $\bar{x} \equiv X$};
{\it ii)}
{\it at $T=0$ the ECP, $F_{ef}(X)$, is equal to the mean energy, 
$E^{min}_0(\bar{x})$ of the MEWP 
whose mean position is at $\bar{x} \equiv X$}. 


In Fig. \ref{f1} we showed the 
probability density $Z(X)^{-1} \rho_X(x)$ obtained from Monte Carlo
path integral simulations of a particle in a double-well potential
at temperature $k_B T = 10^{-3}$ au. This value is a small fraction
of the lowest excitation energy, $\Delta E_0 = 29 \times  10^{-3}$ au 
(tunnel splitting), and therefore a good
approximation to the zero temperature limit. For comparison,
the probability density of MEWP's, obtained numerically as the
ground state of the potential $V_{dp}(x) - f x$, are
shown by broken lines. 
The identity of both probability densities is an important 
result that clarifies the physical meaning of fixed 
centroid path integrals.
In Fig. \ref{f2} the energies $V_{dp}(\bar{x})$ 
and $E^{min}_0(\bar{x})$ are shown.
These curves correspond to the limits of infinite 
and zero temperature for the
ECP, respectively. 

In QTST the second derivative of the ECP, $F_{ef}(X)$ with respect
to $X$, has been shown to be an important quantity to determine
the pre-exponential factor of rate constants. \cite{gillan87}
If the effect of a small external force 
on a stationary quantum state is approximated
by a rigid spatial displacement
of the state, 
it can be shown \cite{ramirez} that {\it in the
zero temperature limit the second derivative of the ECP with respect
to $X$ gives an approximation to the first excitation energy of the 
Hamiltonian $H_f$}, i.e.
 
\begin{equation}
\hbar \left[ \frac{1}{m} \left( \frac{\partial^2 E^{min}_0(\bar{x})}
                  {\partial \bar{x}^2} \right) \right]^{\frac{1}{2}}  
 \simeq \Delta E_f   \,  .
\label{deltaE}
\end{equation}
In Fig. \ref{f3} we display 
the exact values of $\Delta E_f $ 
as a function of $\bar{x}$ (remember
that $f$ is an implicit function of $\bar{x}$). 
The l.h.s of Eq.(\ref{deltaE}), drawn 
by a broken line, was obtained by numerical differentiation
of the function $E^{min}_0(\bar{x})$, shown in Fig. \ref{f2}.
At $\bar{x}=0$ the approximation overestimates the 
energy of the tunnel splitting, $\Delta E_0$, by 25 $\%$.

An approximation to the time evolution of the particle
can be formulated with MEWP's:
{\it i)} only MEWP's with mean position $\bar{x}$ 
and momentum $\bar{p}$ are allowed as time dependent states, i.e.,
these are of the form
$\langle x | \psi_f \rangle \exp ( i \bar{p} x / \hbar ) $.
The value of the mean force 
is $\bar{f}=-f$;         
{\it ii)} the time dependence of $\bar{x}$ and $\bar{p}$ 
is given by the Ehrenfest relations 

\begin{equation}
 \frac{d \bar{x}}{dt} = \frac{\bar{p}}{m}  \,\,; \,\,  
 \frac{d \bar{p}}{dt} = \bar{f}                 \, . 
\label{ehrenfest}
\end{equation}
The total energy of the 
MEWP, $E^{min}_0(\bar{x}) + \bar{p}^2/(2m)$, is conserved 
along this time
evolution. This approximation, formulated without any
reference to path integral concepts, can be recognized as
the zero temperature limit of CMD. The 
formulation of CMD as an approximate 
dynamics based on MEWP's provides new physical insight
into this approximation.
In Fig. \ref{f4} we show a phase space
representation of the time evolution of $\bar{x}(t)$ and $\bar{p}(t)$ for
an initial MEWP with $\bar{x}(0)=-0.7$ au and $\bar{p}(0)=0$ moving in the
potential $V_{dp}$. The exact trajectory
was obtained by numerical solution of the time dependent Sch\"odinger
equation. The CMD trajectory, derived from Eq. (\ref{ehrenfest}), is 
that of a classical particle moving
in the ECP shown in Fig. \ref{f2}.
For comparison the phase space trajectory of a classical particle 
in the potential $V_{dp}$ is also shown. 
The CMD result resembles the real time trajectory.


Summarizing, the centroid density has been derived
in the Schr\"odinger
formulation with the help of a function generating its moments.
The centroid density is related to the response
of the system to a quasi-static force. For
potentials with bound states the centroid density converges
to a delta function as $T \to 0$. 
Several results
have been found in this limit: 
{\it i)} the ECP is the mean energy of
the MEWP's of the potential;
{\it ii)} the second derivative
of the ECP with respect to the centroid position 
approximates the first excitation energy of the system;
{\it iii)} CMD is an
approximate dynamics based on MEWP's. 


This work was supported by DGICYT (Spain) under contract
PB96-0874. We thank E. Artacho and J.J. S\'aenz for helpful 
discussions.


\begin{figure}
\caption{
The continuous lines show normalized probability densities, 
$Z(X)^{-1} \rho_X(x)$, 
obtained by fixed centroid path integral simulations 
at $k_BT = 10^{-3}$ au, 
in the potential $V_{dp}(x)$.
The dotted lines (on top of the continuous ones) are the probability
densities of the ground state of the potential 
$V_{dp}(x)- fx$, that is shown by broken lines for several values of $f$.
}
\label{f1}
\end{figure}
    
\begin{figure}
\caption{
The ECP in the limit $T \to 0$
(broken line) and the double well potential $V_{dp}$  (continuous line).
}
\label{f2}
\end{figure}

\begin{figure}
\caption{
First excitation energy $\Delta E_f$, for the Hamiltonian $H_f$
(continuous line) as a function of the mean position $\bar{x}$
of the ground state of $H_f$,
and its approximation based on the second derivative
of the ECP (broken line).
}
\label{f3}
\end{figure}

\begin{figure}
\caption{
Phase space trajectory showing the time evolution of the
mean position and momentum of an initial state corresponding
to a MEWP of the $V_{dp}$ potential 
with $\bar{x}(0) = -0.7$ au and  $\bar{p}(0)= 0$.
The exact result is compared to
the CMD approximation and to the classical trajectory.
}
\label{f4}
\end{figure}


\begin{references}

\bibitem{ceperley95}
D. M. Ceperley, Rev. Mod. Phys. {\bf 67}, 279 (1995).


\bibitem{cuccoli95}
For a review see: A. Cuccoli, R. Giachetti, V. Tognetti, 
R. Vaia, and P. Verrucchi,
J. Phys.: Condens. Matter {\bf 7}, 7891 (1995).

\bibitem{feynman65}
R.P. Feynman and A.R. Hibbs, {\it Quantum Mechanics and
Path Integrals} (McGraw-Hill, New York, 1965).

 
\bibitem{gillan87}
M. J. Gillan, Phys. Rev. Lett. {\bf 58}, 563 (1987), 
and references therein.

\bibitem{cao94}
J. Cao and G. A. Voth, J. Chem. Phys. {\bf 100}, 5106 (1994).

\bibitem{ramirez}
R. Ram\'{\i}rez, and T. L\'opez-Ciudad, to be published.

\bibitem{acocella95}
D. Acocella, G.K. Horton, and E.R. Cowley, 
Phys. Rev. Lett. {\bf 74}, 4887 (1995).

\bibitem{kleinert93}
H. Kleinert, {\it Pfadintegrale} (Wissenschaftverlag, Mannheim, 1993)
p. 254.

\end{references}
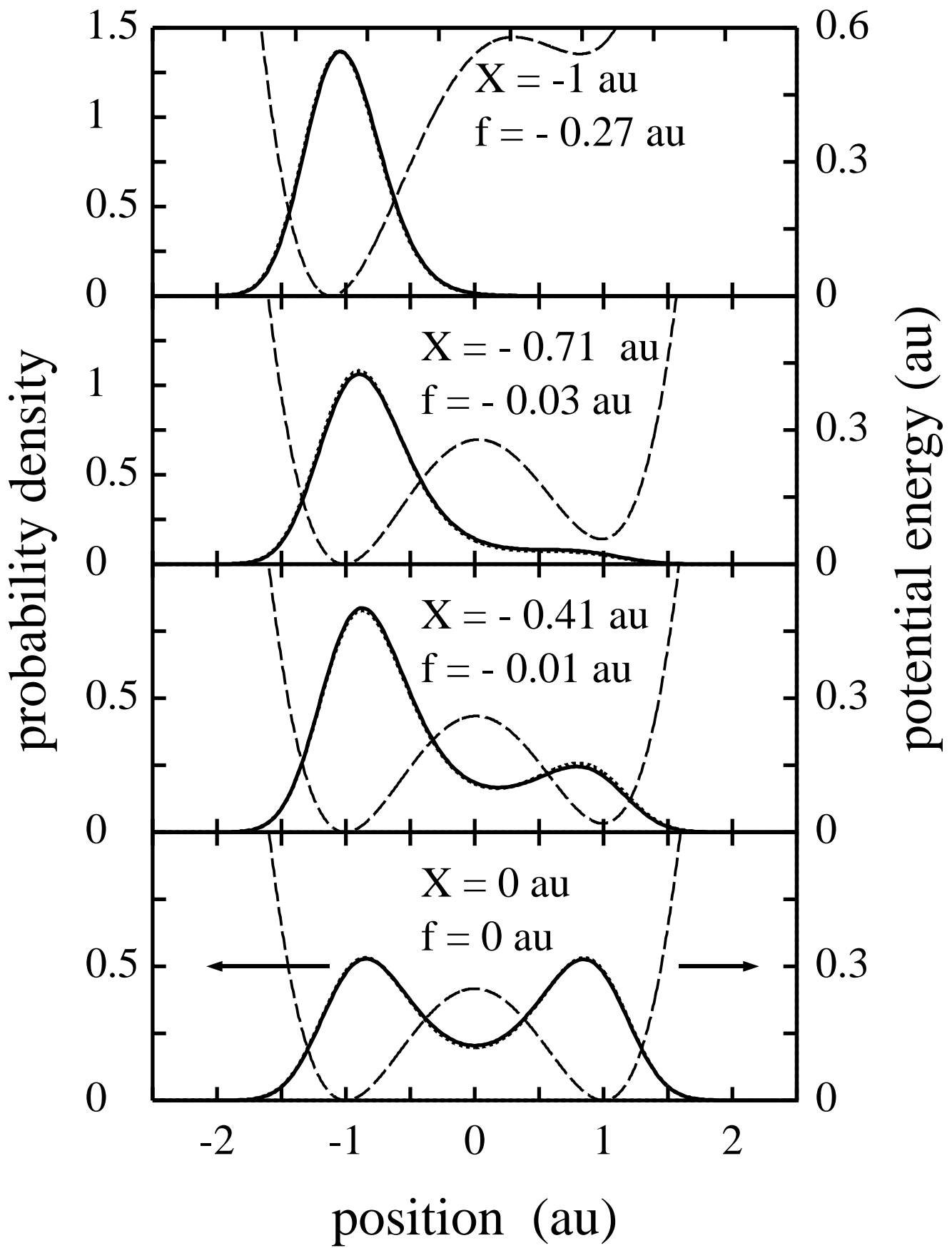
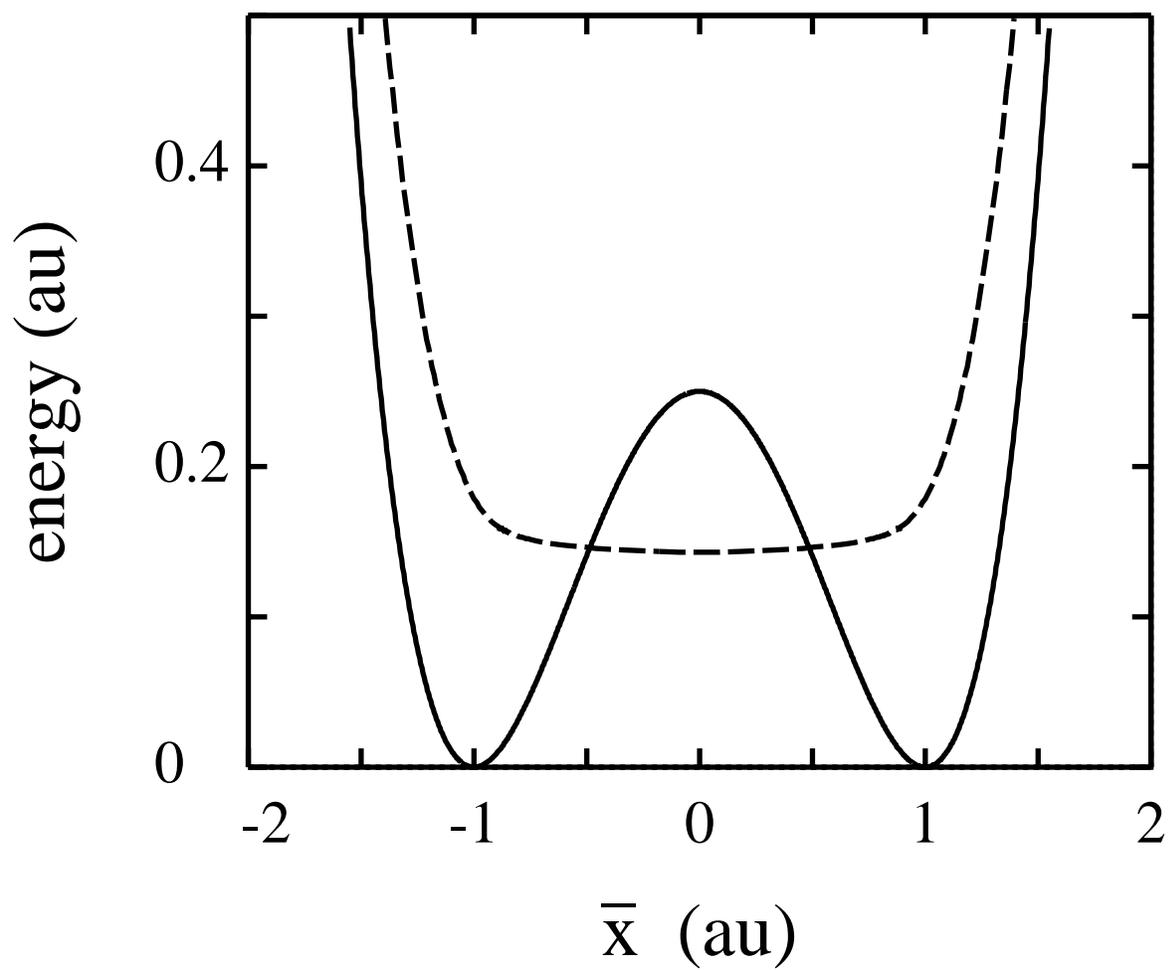
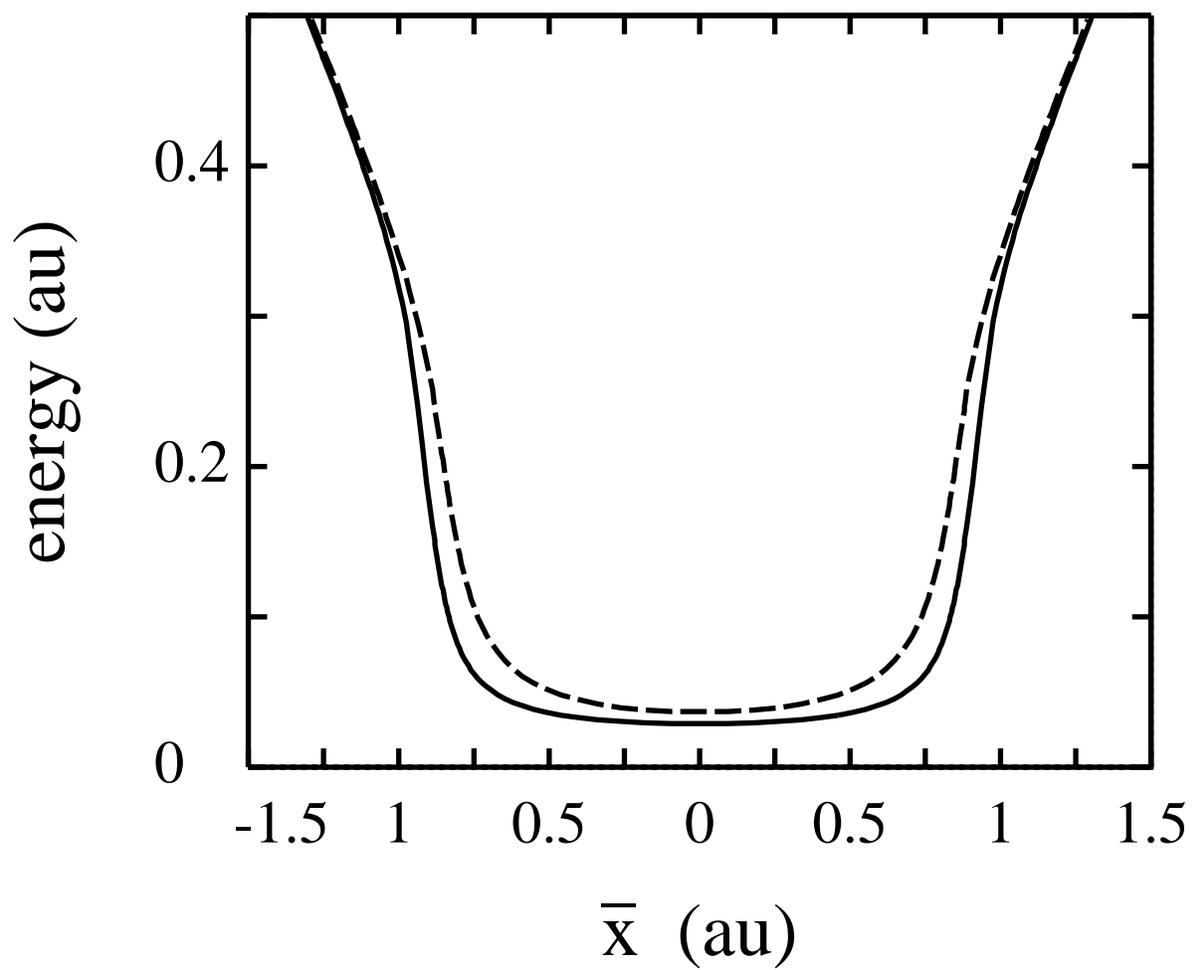
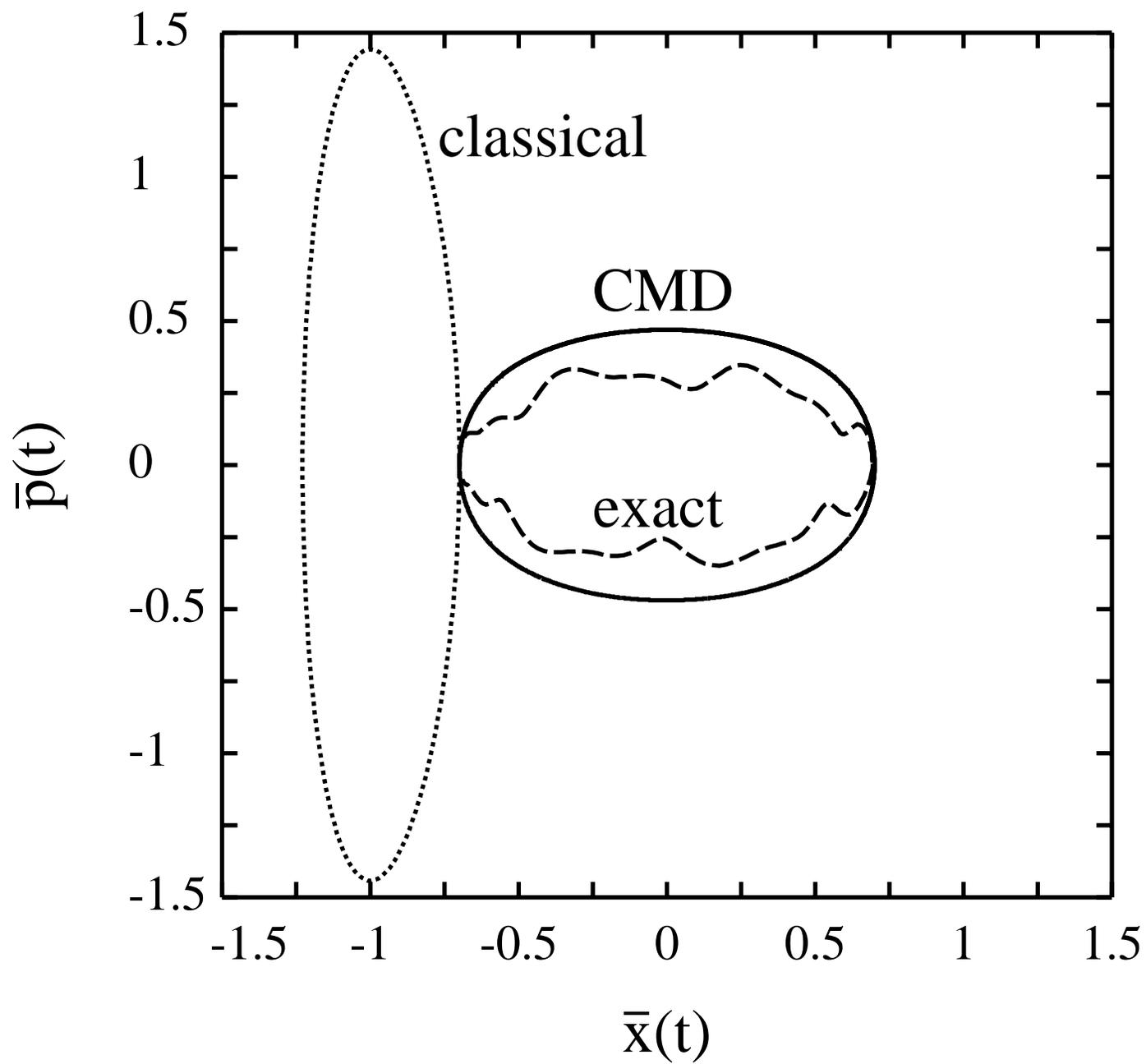
\end{document}